\documentclass[aps,prl]{revtex4}

\usepackage{amssymb,amsmath,graphicx, 
}

\begin{document}

\title{Wavefunction preparation and resampling using a quantum computer}
\author{Alexei Kitaev}
\affiliation{California Institute of Technology, Pasadena, CA, 91125-3600, USA}
\author{William A.\ Webb}
\affiliation{California Institute of Technology, Pasadena, CA, 91125-3600, USA}

\begin{abstract}
We present an algorithm that prepares multidimensional Gaussian wavefunctions on qubit arrays and an application of such wavefunctions to multidimensional resampling, a technique useful in quantum digital simulation.
\end{abstract} 

\maketitle

\section{Introduction}

Shor's algorithms for factoring and discrete logarithms \cite{shor}, as well as Grover's search algorithm \cite{grover} promise substantial quantum speed-up for certain important problems. Another exciting application area for quantum computers is simulation of physical systems, including molecules, strongly correlated spin systems, nuclei, and QCD. The majority of these problems admit a natural description using continuous variables, such as particle coordinates or field modes. (``Mode'' may refer to the field strength at a grid point or to a Fourier component, depending on which representation is chosen.) However, we want to perform the simulation using a \emph{digital} quantum computer, i.e., one composed of qubits. In this paper, we address some general issues pertaining to the discretization of continuous variables and round-off errors in the quantum case. It is worth noting that continuous variables play an important role in algebraic number theory (for example, consider the use of logarithmic coordinates in the proof of Dirichlet's unit theorem \cite{IrelandRosen}.) Thus, we anticipate that discretization techniques might also be helpful in constructing new quantum algorithms for number-theoretic problems.

To see why the discretization is not entirely trivial, consider a quantum particle characterized by a single coordinate $x$. To represent it in a quantum computer, we approximate $x$ by $\tilde{x}=j\delta$, where $j$ is a (binary-encoded) integer and $\delta$ is some small constant (the discretization grid spacing). To simplify the notation, we may rescale $x$ so that $\delta=1$; thus, we approximate real numbers by integers. Now, suppose we want to change this standard grid to a different one, which is equivalent to stretching or squeezing the wavefunction and putting it back on the same grid. We call this transformation \emph{resampling}. If $x$ is mapped to $f(x)$, a continuous wavefunction would transform as follows:
\begin{equation}\label{wfmap}
\psi(x) \mapsto \xi(y)=\bigl(f'(x)\bigr)^{-1/2}\psi(y),\quad
\text{where}\ x = f^{-1}(y).
\end{equation}
In the simplest case, $f(x)$ = $\frac{x}{a}$, so that $x\mapsto\frac{x}{a}$ and $\psi(x)\mapsto\sqrt{a}\psi(ay)$. Naively, this appears to be implementable as a permutation of basis vectors. However, the discrete version of $f$,
\begin{equation}\label{discrete_map}
\tilde{f}(j) = \left\lfloor f(j) \right\rfloor
= \left\lfloor \tfrac{x}{a} \right\rfloor
\end{equation}
is not a permutation! For $a<1$ the transformation is a stretching; it maps integers to a subset of integers, leaving some gaps. For $a>1$ it is a squeezing, therefore two distinct integers may be mapped to the same integer.

Fortunately, transformation~(\ref{discrete_map}) is to be applied to a special class of wavefunction, namely, ones that vary at distances much larger than the grid spacing. In this case, a few lowest bits of $x$ are almost in the uniform superposition and thus can be easily removed or added again while maintaining quantum coherence. But adding or removing $k$ bits corresponds to stretching (resp.\ squeezing) by factor $2^k$. It is easy to generalize this trick to any scaling factor $a$, or to an arbitrary function $f$.

Multidimensional resampling is somewhat more challenging though. Our solution for this case uses multidimensional Gaussian states. Of course, the preparation of such states begins with the one-dimensional case. The transition to higher dimensions requires some care. One may be tempted to say that a general Gaussian is just a product of several one-dimensional Gaussians. However, the implementation of this approach involves the diagonalization of the covariance matrix and a linear change of variables, which in turn requires resampling. To avoid this problem, we use a special linear transformation, namely shearing, and construct its discrete version explicitly. In the rest of the paper, we simply expand this outline.

\section{Preparing a Gaussian in One Dimension}

The states we wish to prepare are discrete approximations to continuous Gaussian functions. In this section we examine these functions and the approximate forms of interest. In one dimension, the Gaussian is just the normal distribution familiar from statistics: $G(x, \sigma, \mu)= e^{-\frac{(x - \mu)^2}{2\sigma^2}}$ (up to a normalization factor). Our algorithm is similar to the uniform superposition preparation technique \cite{ksv} and its generalization to efficiently integrable distributions \cite{grover2}. However, the latter procedures begin with dissecting the distribution at $x_{0}=2^{N-1}$ and setting the highest bit of $x$ according to $\int_{0}^{x_0}|\psi(x)|^{2}\,dx$. In comparison, we first set the \emph{lowest} bit and proceed with generating a discrete Gaussian on either even integers or odd integers. This method is slightly more efficient since it does not involve the Gauss error function, but rather, the Jacobi theta function, which is easier to calculate.

If we treat the Gaussian wavefunction as continuous, it has form 
\begin{equation}
\psi_{\sigma, \mu}(x)
= \frac{1}{\sqrt{\sigma\sqrt{\pi}}} e^{-\frac{(x-\mu)^2}{2\sigma^2}}
\end{equation}
so that 
\begin{equation}
\psi_{\sigma, \mu}^{2}(x)
= \frac{1}{\sigma\sqrt{\pi}} e^{-\frac{(x-\mu)^2}{\sigma^2}}
\end{equation}
and 
\begin{equation}
\int^{\infty}_{-\infty}\psi_{\sigma, \mu}^{2}(x) = 1
\end{equation}
If we consider it to be a function of an integer value, we have 
\begin{equation}
\tilde{\psi}_{\sigma, \mu}(i)
= \frac{1}{\sqrt{f(\sigma, \mu)}} e^{-\frac{(i-\mu)^2}{2\sigma^2}}
\end{equation}
where the function $f$ is related to the third Jacobi theta function
($\theta_{3}$) and is defined as
\begin{equation}
f(\sigma, \mu) = \sum^{\infty}_{n = -\infty}e^{-\frac{(n-\mu)^{2}}{\sigma^{2}}}
\end{equation} 
so that
\begin{equation}
\sum^{\infty}_{i=-\infty}\tilde{\psi}_{\sigma, \mu}^{2}(i) = 1.
\end{equation}
If $\sigma \gg 1$, then $f(\sigma, \mu)$ is approximately equal to
$\int^{\infty}_{-\infty}e^{-\frac{(x-\mu)^2}{\sigma^2}}dx = \sigma\sqrt{\pi}$.
We now define the function
\begin{equation}
\xi_{\sigma, \mu, k}^{2}(i)
= \sum^{\infty}_{j = -\infty}\tilde{\psi}^{2}\bigl(i+j\cdot2^k\bigr)
\end{equation}
Note that this function is periodic with period $2^k$. This allows us to
define the quantum state
\begin{equation}
\left|\xi_{\sigma, \mu, N}\right\rangle
= \sum^{2^N-1}_{i=0}\xi_{\sigma, \mu, N}(i)\left|i\right\rangle
\end{equation}

For $\mu, 2^N-\mu \gg \sigma \gg 1$, this state is very close to a Gaussian with mean $\mu$ and standard deviation $\sigma$. A recursive description of the state,
\begin{equation}
\left|\xi_{\sigma, \mu, N}\right\rangle =
\bigl|\xi_{\frac{\sigma}{2}, \frac{\mu}{2}, N - 1}\bigr\rangle\otimes
\cos\alpha\left|0\right\rangle
+ \bigl|\xi_{\frac{\sigma}{2}, \frac{\mu - 1}{2}, N - 1}\bigr\rangle\otimes
\sin\alpha\left|1\right\rangle,
\end{equation}
where
\begin{equation}\label{alpha}
\alpha = \cos^{-1}\sqrt{f\left(\frac{\sigma}{2},
\frac{\mu}{2}\right)/f(\sigma, \mu)},
\end{equation}
leads almost immediately to an algorithm for preparing it. Note that the sum represented by $f(\sigma, \mu)$ converges, so we can approximate it with a finite number of terms and keep some number of digits of the answer. However, the convergence is poor when $\sigma$ is large. When $\sigma$ is small (which is the case at a late stage of our recursive algorithm), a few terms of the sum will dominate, and only these need be calculated and stored. For large $\sigma$ we use $f(\sigma, \mu)$ = $\sigma\sqrt{\pi}\exp\left(-\frac{\mu^2}{\sigma^2}\right)f\left(\frac{1}{\pi\sigma},\, -i\frac{\mu}{\sigma^{2}\pi}\right)$ to avoid having to sum an exponentially large number of terms.

The algorithm recursively calls the following subroutine, using the input data $(\sigma_{0}, \mu_{0})=(\sigma, \mu)$ and some number of supplementary qubits. Note that the $N$ qubits are indexed 0 to $N-1$, where the 0-th qubit corresponds to the rightmost qubit in a binary representation.


\begin{enumerate}
\item 
\begin{itemize}
\item[(a)] Calculate and store $\alpha$ (see eqn. $\eqref{alpha}$ ) and $\alpha/2\pi$ in a k-bit binary representation on supplementary qubits.
\item[(b)] Apply the rotation operator R($\alpha$) to the 0-th qubit
\begin{equation}
R(\alpha) = \left[
\begin {array}{cc}
\cos\alpha & -\sin\alpha\\
\noalign{\medskip}
\sin\alpha & \cos\alpha
\end {array}
\right]
\end{equation}
using the bits of $\alpha/2\pi$ as described below.
\item[(c)] Uncompute $\alpha$ and $\alpha/2\pi$.
\end{itemize}

\item Compute $(\sigma_{1}, \mu_{1})$, where $\sigma_{1}$=$\sigma_{0}/2$ and $\mu_{1}$ = $\mu_{0}/2$ if the previously rotated qubit is $\left|0\right\rangle$ and $\mu_{1}$ = $(\mu_{0}-1)/2$ if it is $\left|1\right\rangle$. Note that all pairs $(\sigma_{i}, \mu_{i})$ are stored in supplementary memory until a cleanup phase, see below.

\item On the remaining $N-1$ qubits prepare state $\left|\xi_{\sigma_{1}, \mu_{1}, N-1}\right\rangle$.
\end{enumerate}


Note that for the last ((N-1)th) qubit, we proceed only through step 1 (c), as after this qubit is rotated we do not need another pair of parameters. After this last qubit is rotated, right before termination, a last series of steps is performed in which all stored pairs $(\sigma_{i}, \mu_{i})$ are cleared. To do this, the machine erases $(\sigma_{N-1}, \mu_{N-1})$ by recalculating it using $(\sigma_{N-2}, \mu_{N-2})$ and the last output qubit, performing a second controlled XOR to the region where $(\sigma_{N-1}, \mu_{N-1})$ is stored, thus clearing $(\sigma_{N-1}, \mu_{N-1})$ from memory. The process is repeated on $(\sigma_{N-2}, \mu_{N-2})$, $(\sigma_{N-3}, \mu_{N-3})$ and so on until the supplementary memory is cleared.

Since quantum recursion can be counterintuitive and difficult to describe verbally, this algorithm is perhaps easier to understand when presented in circuit form. The figure shows the gist of the algorithm as well as some of the workings of the supplementary qubits.  Note that the double line across the top denotes the classical part of the computer, which in this scheme functions only to transfer the specifications of the Gaussian to the quantum part at the end of the process to clear memory. The single lines with forward slashes (/) overlying their left ends represent multiple qubits in the supplementary memory. Note that each such line represents the same array at different stages of the computation. We deviate slightly from conventional notation by introducing the downward arrow conditional, which represents a generalized form of controlled operation.

The downward arrow terminating on the XOR symbol on the supplementary array directly above the $i$th qubit denotes that the previously used pair $(\sigma_{i-1},\mu_{i-1})$ and the $(i-1)$th qubit (which was just rotated) are used to control the calculation of $(\sigma_{i},\mu_{i})$ and $\alpha_{i}$. Note that classically, $\alpha_{i}$ is a rotation angle equal to $\sqrt{f(\sigma_{i-1}/2, \mu_{i-1}/2)/f(\sigma_{i-1}, \mu_{i-1})}$ if the $(i-1)$-th qubit is $\left|0\right\rangle$ and $\sqrt{f(\sigma_{i-1}/2, (\mu_{i-1}-1)/2)/f(\sigma_{i-1}, \mu_{i-1})}$ if the $(i-1)$-th qubit is $\left|1\right\rangle$. However, since the previous qubit is always in a superposition state, $\alpha_{i}$ is always entangled with the previous qubits.

The downward arrow terminating on the i-th rotation operator box indicates that the rotation angle is specified by the controlling qubits, specifically, the qubits in the supplementary containing $\alpha_{i}$ (which is shown directly to the right of the downward arrow). This rotation is actually a series of standardized rotations, each controlled by a single qubit of the register storing $\alpha_{i}$, as described below.

\begin{figure}[htbp]   
\includegraphics[width=14cm]{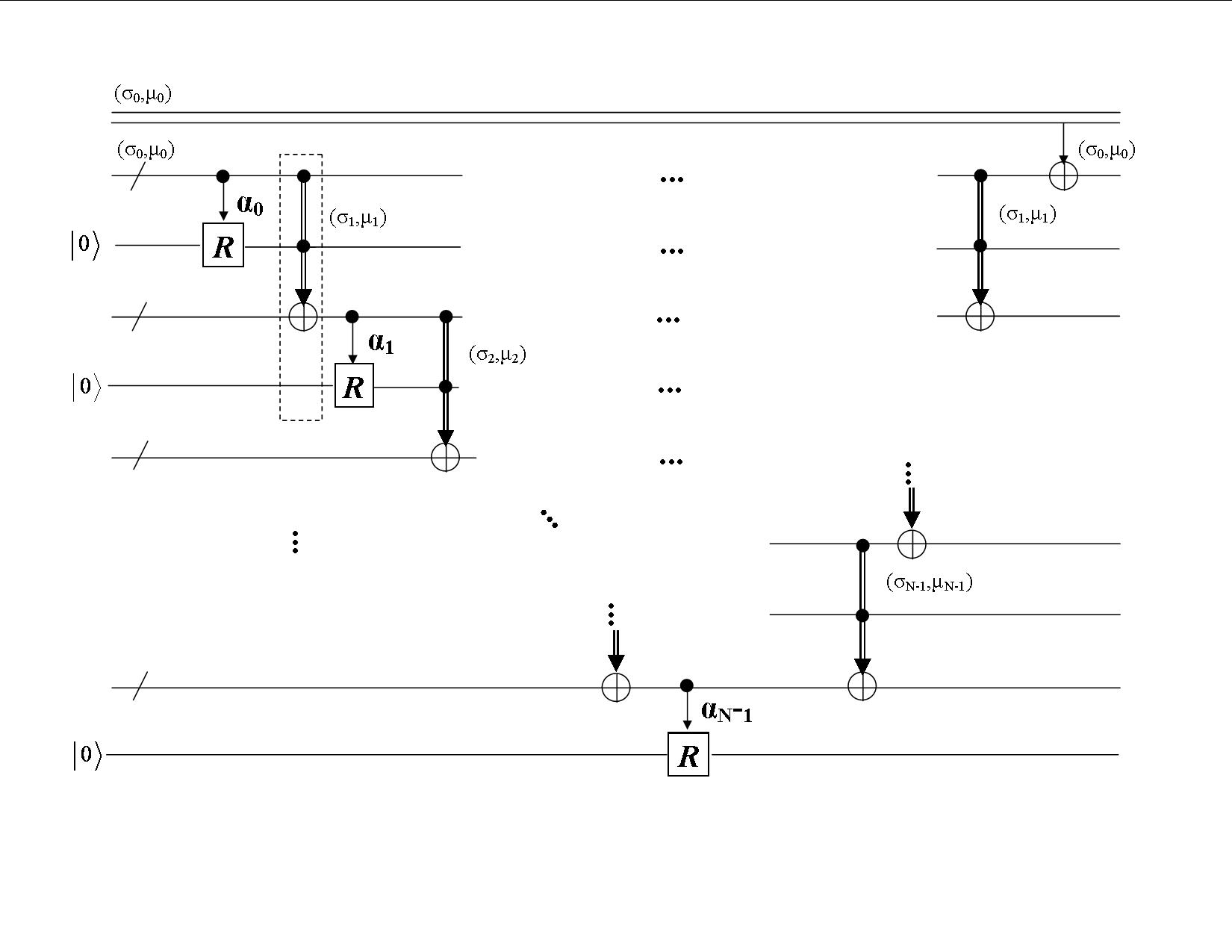}
\includegraphics[width=14cm]{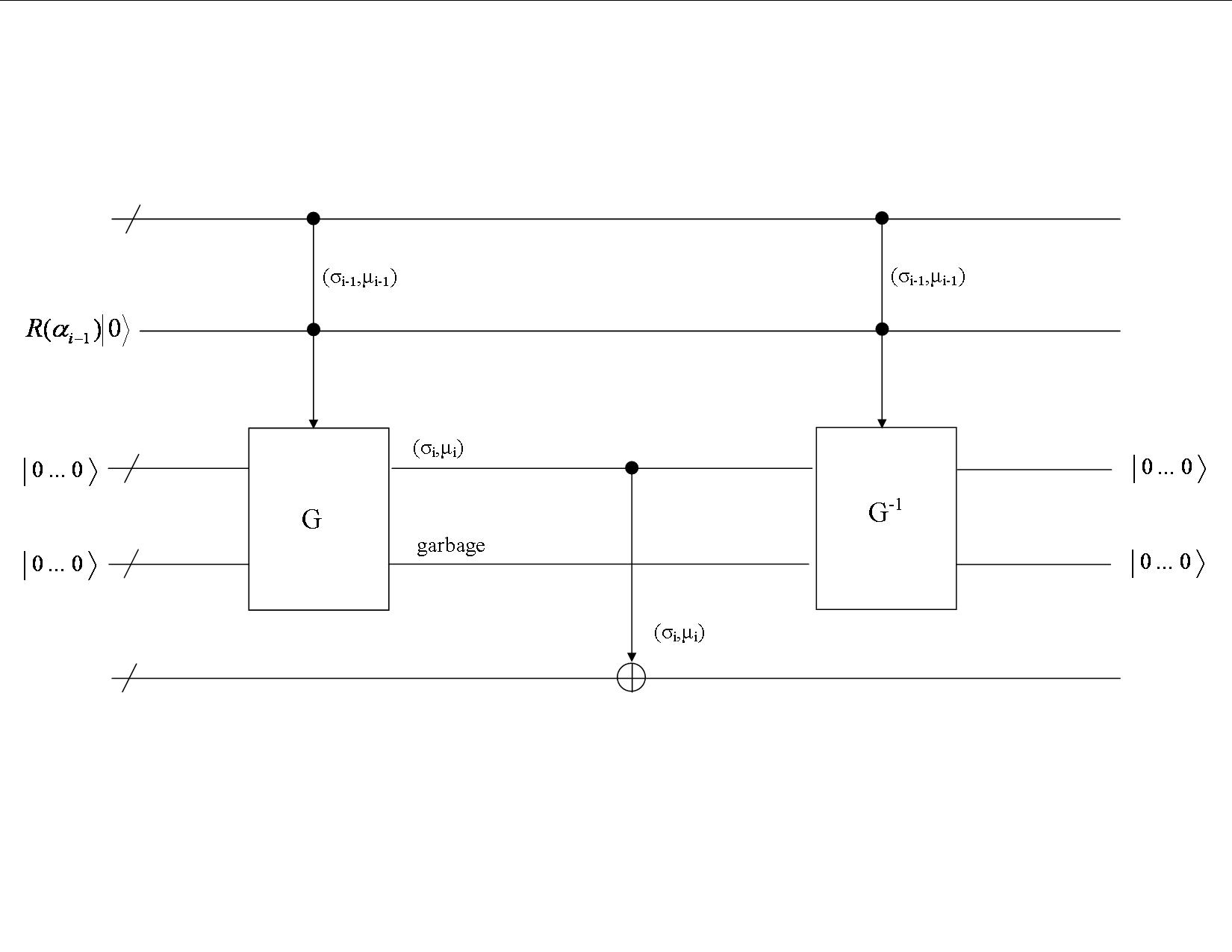}
\vspace{-1cm}   
\caption{Schematic circuit for preparing a Gaussian wavefunction. The top part of the figure shows the general scheme of the recursion. The lower figure is a more detailed view of what occurs in a representative block of the process like the boxed section of the circuit above. The top and bottom multi-qubit wires represent different regions of the semipermanent supplementary memory where all pairs $(\sigma_{i}, \mu_{i})$ are stored until the end of computation. The multi-qubit wires feeding into the black box G are the ``working" supplementary memory that is inverted and cleared in each step.}
\end{figure}

Due to finite computational resources, the rotation operator must be implemented approximately. Note that since $\alpha$ is a rotation angle between 0 and $2\pi$, so $0 \leq \alpha/2\pi\leq 1$ and we can approximate in $k$-digits of binary notation that $\alpha/2\pi \approx \sum^{k}_{i=1}\frac{a_i}{2^{i}}$ so that $\alpha \approx \sum^{k}_{i=1}\pi\frac{a_i}{2^{i-1}}$. Note that the $a_{i}$ are either 1 or 0, as this is a binary representation. The rotation operator is realized as a sequence of at most $k$ standard rotations as $R(\alpha)\approx R(\pi/2^{k-1})^{a_{k}}\cdots R(\pi/2)^{a_{2}}R(\pi)^{a_{1}}$. The precision of this operation on the entire array (assuming the standard rotations are exact) is thus $\delta = O(N2^{-k})$ and the size of the circuit required to execute the algorithm is polynomial in $n + \log(1/\delta)$.

\section{Preparing a Gaussian in multiple dimensions} 
In $S$ dimensions, the general Gaussian function is $G(\vec{x}) = Ce^{-\vec{x}^{T}A\vec{x}/2}$, where $C$ is a constant, $A$ is a real symmetric positive definite matrix ($A^{T}$=$A$, and $\vec{x}^{T}A\vec{x} \geq 0$ $\forall x \neq 0$), and $\vec{x}$ is vector with S components, $\vec{x} = \vec{x'} - \vec{\mu}$ where $\vec{x'}$ denotes position in $S$ space and $\vec{\mu}$ is a vector denoting the mean of the distribution.

We assume that the register (call it $R$) contains $Sk$ qubits, where $k$ is some integer. Thus, we can break the array into $S$ subarrays (index them $R_{i}$ for $1 \leq i \leq S$) each containing $k$ qubits, and write $R$=$R_{1}R_{2}...R_{S}$, with multiplication representing array concatenation. The state space corresponding to $R$ is of dimension $2^{Sk}$. The wavefunction we wish to prepare has form
\begin{equation}
\left|\psi\right\rangle
= C\sum_{\vec{n}\in{B^{S}}}e^{-\vec{n}^{T}A\vec{n}/2}\left|\vec{n}\right\rangle
\end{equation}
In this expressions $C^2$ = $\pi^{-S/2}\sqrt{\det{A}}$ and $\vec{n} \in B^{S}$, $B$ = $[-2^{k-1},..., 2^{k-1}-1]$, where negative integers are represented using two's complement. In the quantum computer, $\left|n\right\rangle$ is the basis vector formed by concatenating all binary representations of the entries of $\vec{n}$. For example, if $n=9$, in binary representation, $n=1001$, so if we assume $k$=2, $\vec{n}$ = $(10, 01)^{T}$ = $(2,1)^{T}$. For $R_{i}$, the basis state $\left|m\right\rangle$ corresponds to the integer $m$ along the $i$th coordinate axis in $S$ space. Thus, each subarray generates one coordinate axis of the space in question. Note that this space is not a state space in the system, rather, each integer point in the cube described corresponds to one dimension/basis vector of the system's state space. The reason for the distinction will be clear momentarily.

Our method of preparing the general $S$-dimensional Gaussian relies on a special case of the problem. Note that if the matrix $A$ in $e^{-x^{T}Dx/2} = Ce^{-\vec{x}^{T}A\vec{x}/2}$ is diagonal (in which case, we call it $D$), then $G(\vec{x})$ factors into 1-dimensional Gaussians as $G(\vec{x})= \prod^{N}_{i=1}\exp(-d_{i}x^{2}_{i}/2)$ where $d_{i}$ is the $i$th diagonal element of $D$ and $x_{i}$ the $i$-th entry of $\vec{x}$.

Thus in this special case, the state to prepare becomes (with $\vec{n}$=$(n_{1},...,n_{S})$, where the $n_{i}$ are $k$-digit binary integers)

\begin{equation}
\left|\psi\right\rangle =
C\sum_{\vec{n}\in{B^{S}}}e^{-\vec{n}^{T}D\vec{n}/2}\left|\vec{n}\right\rangle
= C\sum_{\vec{n}\in{B^S}}
\prod^{S}_{i=1}\exp\left(-\tfrac{1}{2}d_{i}n^{2}_{i}\right)
\left|\vec{n}\right\rangle
= C\bigotimes^{S}_{i=1}\left(\sum_{n\in B}
\exp\left(-\tfrac{1}{2}d_{i}n^{2}\right)\left|n\right\rangle\right)
\end{equation}

This state is just a product of 1-dimensional Gaussians, which we have already shown how to prepare. We can prepare this state by preparing $N$ simple Gaussians on $k$ qubit subarrays. The overall normalization constant (here denoted by $C$) is accounted for by the normalization constants of the Gaussians on subarrays.
 
Some simple properties of linear algebra and a change of variables allow us to transform this easily prepared state into any arbitrary Gaussian. Any real symmetric positive definite matrix $A$ can be decomposed as $A$ = $(M^{T})^{-1}D(M^{-1})$ for some $D$ and $M$, where $D$ is diagonal and $M$ is upper triangular with 1's on the diagonal (in other words, a multidimensional shearing matrix). Thus, if we apply transformation $\vec{n} \rightarrow \vec{x}=M\vec{n} \Leftrightarrow \vec{n} = M^{-1}\vec{x}$, then
\begin{equation}
\exp(-\vec{n}^{T}D\vec{n}/2)
= \exp\left(-\frac{\vec{x}^{T}(M^{T})^{-1}DM^{-1}\vec{x}}{2}\right)
= \exp\left(-\frac{\vec{x}^{T}A\vec{x}}{2}\right)
\end{equation}

The matrix $M$ can be decomposed as the product of shearing matrices with exactly one non-zero off-diagonal entry each. Note that $M$ acts numerically on the basis vectors in a classically reversible fashion (provided relevant parameters are stored). It is not a unitary operator and is not applied directly to the state space of the system. Rather, it maps a vector $\vec{n}$ to a vector $\vec{x}$ (and the corresponding basis vector $\left|n\right\rangle$ to $\left|x\right\rangle$) as in the following example.

Let $M$ be a square, 2 by 2 upper triangular matrix with 1's the diagonal and a real number $\alpha$ in the upper right entry and let $\vec{n}$ = $\left(n_{1}, n_{2}\right)^{T}$. Then, we see that

\begin{equation}
M = \left(
\begin {array}{cc}
1 & \alpha \\
\noalign{\medskip}
0 & 1
\end {array}
\right), \hspace{.5cm}
\left(
\begin {array}{c}
x_{1}\\
\noalign{\medskip}
x_{2}
\end {array}
\right)=
M\vec{n} =
\left(
\begin {array}{cc}
1 & \alpha \\
\noalign{\medskip}
0 & 1
\end {array}
\right)
\left(
\begin {array}{c}
n_{1}\\
\noalign{\medskip}
n_{2}
\end {array}
\right) =
\left(
\begin {array}{c}
n_{1}+ \alpha{n_{2}} \\
\noalign{\medskip}
n_{2}
\end {array}
\right)
\approx
\left(
\begin {array}{c}
n_{1} + \left\lfloor \alpha{n_{2}}\right\rfloor \\
\noalign{\medskip}
n_{2}
\end {array}
\right)
\end{equation}

\begin{equation}
\left(
\begin {array}{c}
n_{1} \\
\noalign{\medskip}
n_{2}
\end {array}
\right)
=
\left(
\begin {array}{c}
x_{1} - \left\lfloor \alpha{x_{2}}\right\rfloor \\
\noalign{\medskip}
x_{2}
\end {array}
\right)
\end{equation}

Thus, $M$ can act on basis vectors numerically by matrix multiplication with rounding, which, provided the specifications for $M$ are stored, is an efficiently reversible classical operation that can thus be carried out on a quantum computer. Applying $M$ to a superposition of states results in a changes of variable of the kind described above, which turns our set of independent Gaussians into the general Gaussian,
\begin{equation}
C\sum_{\vec{n}\in{B^{S}}}e^{-\vec{n}^{T}D\vec{n}/2}\left|\vec{n}\right\rangle
\rightarrow
C\sum_{\vec{x}\in{B^{S}}}e^{-\vec{x}^{T}A\vec{x}/2}\left|\vec{x}\right\rangle.
\end{equation}
Thus, we have an algorithm for preparing the general Gaussian function in $N$
dimensions.
 
\section{Resampling}

One application of the Gaussian states described above is multidimensional resampling. This is more clear after a description of resampling in one dimension. The goal is to stretch or contract a wavefunction with respect to some function $f$, see eqn.~(\ref{wfmap}). (We assume that $f$ is smooth, monotone, and its derivative is bounded from above and from below.) Such deformations are easily accomplished, provided that $\psi$ is a slowly varying function ($\psi(x)$ and $\psi(x\pm 1)$ are very close for all $x$ in the range of interest).  Assume we have two qubit ensembles, $A$ and $B$, each containing $N$ qubits. The basis states of ensemble $A$ correspond to the $x$ axis, those of ensemble $B$ to the $y$ axis. The desired original wavefunction $\psi$ is prepared on ensemble $A$, while ensemble $B$ is prepared in state $\left|0^N\right\rangle$. The state of the combined system is thus
\begin{equation}
\left|\eta_{0}\right\rangle = \sum^{2^N-1}_{x=0}\psi(x)\left|x,0^N\right\rangle
\end{equation}
If we want to perform a resampling with respect to $f(x) = \left\lfloor \frac{x}{a}\right\rfloor$, then we prepare the uniform superposition state
\begin{equation}
\left|\kappa_{n}\right\rangle
= \frac{1}{\sqrt{2n}}\sum^{n-1}_{j=-n}\left|j\right\rangle
\end{equation}
on ensemble B using an algorithm described in Problem 9.4 of \cite{ksv}. We describe how parameter $n$ is selected shortly. After this operation, the whole ensemble is in state
\begin{equation}
\left|\eta_{1}\right\rangle
= \frac{1}{\sqrt{2n}}\sum^{2^N-1}_{x=0}\psi(x)
\sum^{n-1}_{j=-n}\left|x,j\right\rangle
\end{equation}
A simple reversible computation can efficiently transform this state into
\begin{equation}
\left|\eta_{2}\right\rangle
= \frac{1}{\sqrt{2n}}\sum^{2^N-1}_{x=0}\psi(x)
\sum^{n-1}_{j=-n}\left|x,\left\lfloor \frac{x}{a}\right\rfloor + j\right\rangle
\end{equation}
Recall that we wish to prepare the state $\xi(y) = \sqrt{a}\psi\left(f(ay)\right)$ on an $N$-qubit ensemble. Note that if we started with that wavefunction prepared on ensemble $B$ with ensemble $A$ initialized to $\left|0^N\right\rangle$, we could carry out a transformation analogous to that above to yield the state
\begin{equation}
\left|\tilde{\eta}_{2}\right\rangle
= \frac{1}{\sqrt{2n}}\sum^{2^N-1}_{y=0}\psi(ay)
\sum^{an-1}_{j=-an}\bigl|\left\lfloor ay\right\rfloor+j, y\bigr\rangle.
\end{equation} 
If we think of $A$ and $B$ as generating coordinate axes $x$ and $y$, then wavefunctions can be visualized as sets of points in the $xy$ plane, specifically, as points occupying the vertices of the infinite grid with integer spacing. In this scheme, $\left|\eta_{2}\right\rangle$ and $\left|\tilde{\eta}_{2}\right\rangle$ appear as bands of width $2n$ about the line $y = \frac{x}{a}$. The amplitudes of basis vectors in $|\eta_{2}\rangle$ and $|\tilde{\eta}_{2}\rangle$ agree on that line; the only difference is that the amplitudes in $|\eta_{2}\rangle$ are constant on vertical strips whereas the amplitudes in $|\tilde{\eta}_{2}\rangle$ are constant on horizontal strips. If $n$ is much larger than the grid spacing but small enough relative to the grid length at which $\psi$ and $\xi$ vary, the amplitudes in intersecting horizontal and vertical strips are close. Thus we have the approximate equality $\left|\eta_{2}\right\rangle \cong \left|\tilde{\eta}_{2}\right\rangle$. This means that by choosing $N$ and $n$ appropriately and preparing the state $\left|\eta_{2}\right\rangle$, it is virtually equivalent to having prepared the state $\left|\tilde{\eta}_{2}\right\rangle$. If we then apply (to the combined ensemble) the inverse of the computation we would have used to prepare $\left|\tilde{\eta}_{2}\right\rangle$, it is as if we prepared and then uncomputed $\left|\tilde{\eta}_{2}\right\rangle$, leaving the combined ensemble in (or rather, arbitrarily close to) the state such that the wavefunction on $B$ is $\xi(y)$ and the wavefunction on $A$ is $\left|0^N\right\rangle$. Thus if we can prepare $\psi(x)$ satisfying the necessary smoothness conditions, we can use resampling with supplementary qubits to transform the state to one arbitrarily close to $\xi(y)$.

\begin{figure}[htbp]   
\includegraphics[width=14cm]{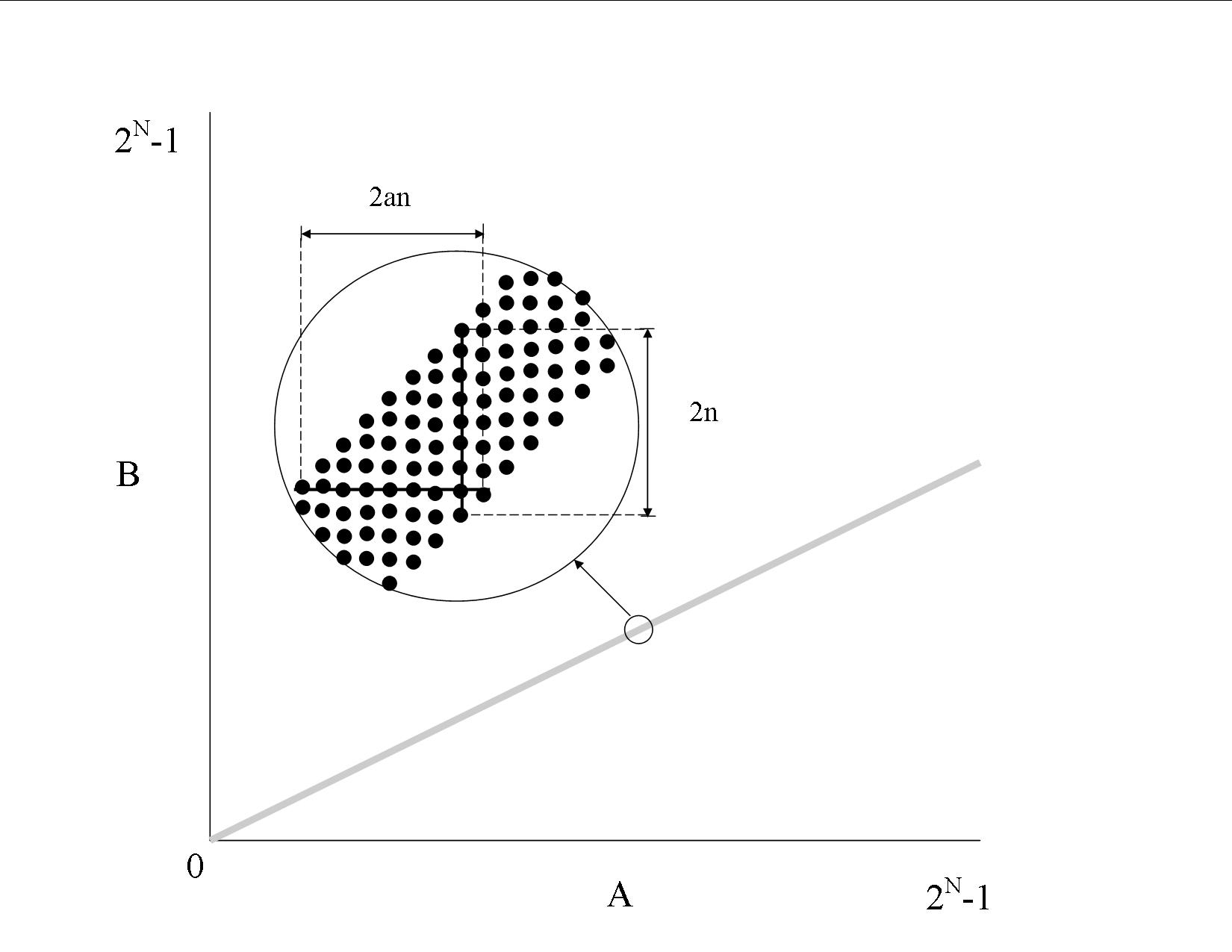}
\vspace{-0.5cm}   
\caption{Conceptual representation of states $\left|\tilde{\eta}_{2}\right\rangle$ and $\left|\eta_{2}\right\rangle$. The grey region (and the dotted region in the inset) represents the basis states of the combined ensemble AB that have non-zero amplitudes. In state $\left|\eta_{2}\right\rangle$ the amplitudes on vertical strips are equal, in state $\left|\tilde{\eta}_{2}\right\rangle$ amplitudes are constant on horizontal strips. If n is much larger than the spacing between basis vectors but much smaller than the size of the grid, the amplitudes of basis vectors in a vertical strip will be roughly equal to those of the basis vectors of any horizontal strip that intersects it. When this is the case, $\left|\tilde{\eta}_{2}\right\rangle \cong \left|\eta_{2}\right\rangle$ and resampling will work.}
\end{figure}

This particular transformation takes advantage of the fact that uniform superpositions on an interval are invariant under stretching or contracting transformation. Multidimensional resampling requires a family of states that is similarly invariant under a general linear transformation. Multidimensional Gaussian states satisfy this condition and thus may be used for multidimensional resampling.

We illustrate with a simple example. Note that in the procedure above, if we prepare on ensemble $B$ the state $\left|\xi_{\sigma, \mu, N}\right\rangle$ with parameters $\sigma$ and $\mu$ such that
\begin{equation}
\left|\xi_{\sigma, \mu, N}\right\rangle
\cong \sum^{2^N-1}_{i=0}e^{-\frac{(i-\mu)^2}{2\sigma^2}}\left|i\right\rangle
\cong \bigl|\xi_{\sigma, \mu}\bigr\rangle
= \sum^{\infty}_{i=-\infty}e^{-\frac{(i-\mu)^2}{2\sigma^2}}\left|i\right\rangle
\end{equation}
with $\mu \gg \sigma \gg 1$ and $\sigma \cong 2n$ instead of the uniform superposition state $\left|\kappa_{n}\right\rangle$, the procedure still works. Intuitively, we prepare a series of vertical, parallel Gaussian-weighted strips centered somewhere above the $x$ axis and shift them up along the line $y$ = $\frac{x}{a}$ so that each strip is centered at the line. The result is that the entire grid has non-zero amplitudes, but all amplitudes except those immediately about $y$ = $\frac{x}{a}$ are very, very small, so the result looks very similar to Figure~2 and is amenable to resampling as described.

\section{Acknowledgements}

This research was funded by Caltech's Summer Undergraduate Research Fellowship (SURF) Program. W.W. would like to thank Ning Bao, Erik Winfree, and David Renshaw for helpful comments and discussion.


\end{document}